\documentstyle[12pt]{article}
\newcommand{\bra}{\langle}
\newcommand{\ket}{\rangle}
\newcommand{\R}{\mbox{\boldmath $ R $}}
\newcommand{\Z}{\mbox{\boldmath $ Z $}}
\newcommand{\map}{\mbox{Map}}
\newcommand{\diff}{\mbox{Diff}}

\makeatletter
\@addtoreset{equation}{section}
\makeatother

\begin{document}
\baselineskip 7mm
\begin{flushright}
DPNU-94-35
\\
hep-th/9408096
\\
August 1994
\end{flushright}
\baselineskip 10mm
\begin{center}
{\LARGE \bf %
Topology and quantization \\
of abelian sigma model \\ in $ (1 + 1) $ dimensions}\footnote{%
Contributed to Yamada Conference
(XXth International Colloquium on Group Theoretical Methods in Physics;
July 1994 at Toyonaka in Japan)}
\vspace{12mm}
\\
{\Large Shogo Tanimura}\footnote{%
e-mail address : tanimura@eken.phys.nagoya-u.ac.jp}
\vspace{6mm}
\\
{\large \it Department of Physics, Nagoya University, \\
Nagoya 464-01, Japan}
\vspace{16mm}
\\
\begin{minipage}[t]{120mm}
\baselineskip 6mm
It is known that there exist an infinite number of inequivalent quantizations
on a topologically nontrivial manifold
even if it is a finite-dimensional manifold.
In this paper we consider the abelian sigma model in $ (1+1) $ dimensions
to explore a system having infinite degrees of freedom.
The model has a field variable $ \phi : S^1 \to S^1 $.
An algebra of the quantum field is defined
respecting the topological aspect of this model.
A central extension of the algebra is also introduced.
It is shown that
there exist an infinite number of unitary inequivalent representations,
which are characterized by a central extension and a continuous parameter
$ \alpha $ $ ( 0 \le \alpha < 1 ) $.
When the central extension exists,
the winding operator and the zero-mode momentum obey a nontrivial commutator.
\end{minipage}
\end{center}
%
\newpage
\baselineskip 7mm
\section{Introduction}
In both field theory and string theory
there are several models which have manifold-valued variables.
 For instance, the nonlinear sigma model has a field variable
$ \phi : \R^{\, 3} \to G/H $, where $ G/H $ is a homogeneous space.
This manifold $ G/H $ is closely related
to vacua associated with spontaneous symmetry breaking.
As another example,
the toroidal compactification model of closed bosonic string has a variable
$ X : S^1 \to T^{\, n} $.
To study global aspects of these models in quantum theory,
we should have a quantization scheme respecting topological nature.
However in the scheme of usual canonical quantization and perturbation method,
the global aspects are obscure.
\par
On the other hand it is known \cite{OK}, \cite{Tani2} that
there exist an infinite number of inequivalent quantizations
on a topologically nontrivial manifold
even if it is a finite-dimensional manifold.
Unfortunately
it remains difficult
to extend those quantization schemes to include field theory.
\par
In this paper we consider the abelian sigma model in $ (1+1) $ dimensions
to explore a system having infinite degrees of freedom.
In the context of classical theory,
a field variable of the model is a map from $ S^1 $ to $ S^1$.
An algebra of the quantum field is defined
respecting the topological aspect of this model.
Special attention is paid for the zero-mode and the winding number.
A central extension of the algebra is also introduced.
Representation spaces of the algebra are constructed
using the Ohnuki-Kitakado representation and the Fock representation.
It is shown that
there exist an infinite number of unitary inequivalent representations,
which are parametrized by a continuous parameter
$ \alpha $ $ ( 0 \le \alpha < 1 ) $.
It is expected that this model gives a physical insight to
nonlinear sigma models of field theory
and orbifold models of string theory.
%
\section{Ohnuki-Kitakado representation}
Here we briefly review quantum mechanics of a particle on $ S^1 $
considered by Ohnuki and Kitakado~\cite{OK}.
They assume that
a unitary operator $ \hat{U} $ and a self-adjoint operator $ \hat{P} $
satisfy the commutation relation
\begin{equation}
	[ \, \hat{P} \, , \, \hat{U} \, ] = \hat{U}.
	\label{2.1}
\end{equation}
An irreducible representation of the above algebra is defined to be
quantum mechanics on $ S^1 $.
The operators $ \hat{U} $ and $ \hat{P} $ are called
a position operator and a momentum operator, respectively.
It is shown below that this naming is reasonable.
\par
A representation space is constructed as follows.
Let $ | \alpha \ket $ be an eigenvector of $ \hat{P} $
with a real eigenvalue $ \alpha $;
$ \hat{P} \, | \alpha \ket = \alpha \, | \alpha \ket $.
Assume that $ \bra \alpha | \alpha \ket = 1 $.
The commutator (\ref{2.1}) implies that
the operator $ \hat{U} $ increases an eigenvalue of $ \hat{P} $ by a unit.
If we put
\begin{equation}
	| n + \alpha \ket := \hat{U}^n \, | \alpha \ket
	\;\;\;
	( n = 0 , \pm 1 , \pm 2 , \cdots ),
	\label{2.2}
\end{equation}
it is easily seen that
\begin{eqnarray}
	&&
	\hat{P} \, | n + \alpha \ket = ( n + \alpha ) \, | n + \alpha \ket,
	\label{2.3}
	\\
	&&
	\hat{U} \, | n + \alpha \ket = | n + 1 + \alpha \ket.
	\label{2.4}
\end{eqnarray}
Unitarity of $ \hat{U} $ and self-adjointness of $ \hat{P} $ imply that
\begin{equation}
	\bra m + \alpha | n + \alpha \ket = \delta_{ m \, n }.
	\label{2.5}
\end{equation}
Let $ H_\alpha $ denote the Hilbert space defined by completing
the space of finite linear combinations of
$ | n + \alpha \ket \, ( n = 0 , \pm 1 , \pm 2 , \cdots ) $.
By (\ref{2.3}) and (\ref{2.4}), $ H_\alpha $ becomes
an irreducible representation space of the algebra (\ref{2.1}).
\par
$ H_\alpha $ and $ H_\beta $ are unitary equivalent
if and only if the difference $ ( \alpha - \beta ) $ is an integer.
Consequently there exists an inequivalent representation
for each value of the parameter $ \alpha $ ranging over $ 0 \le \alpha < 1 $.
At this point, quantum mechanics on $ S^1 $ is in contrast to
quantum mechanics on $ \R $.
 For the one on $ \R $, it is well-known that
the algebra of the canonical commutation relations has a unique irreducible
representation upto unitary equivalence.
\par
To clarify the physical meaning of the parameter $ \alpha $,
they \cite{OK} study eigenstates of the position operator $ \hat{U} $.
If we put
\begin{equation}
	| \lambda \ket
	:=
	\sum_{n \, = \, - \infty}^{\infty} \,
	e^{ - i \, n \, \lambda } \, | n + \alpha \ket,
	\label{2.6}
\end{equation}
it follows that
\begin{eqnarray}
	&&
	\hat{U} \, | \lambda \ket = e^{ i \, \lambda } \, | \lambda \ket,
	\label{2.7}
	\\
	&&
	| \lambda + 2 \pi \ket = | \lambda \ket,
	\label{2.8}
	\\
	&&
	\bra \lambda | \lambda' \ket =
	2 \pi \, \delta( \lambda - \lambda' ).
	\label{2.9}
\end{eqnarray}
In the last equation
it is assumed that
the $ \delta $-function is periodic with periodicity $ 2 \pi $.
It is also easily seen that
\begin{equation}
	\exp( - i \mu \hat{P} ) | \lambda \ket
	=
	e^{ -i \, \alpha \, \mu } \, | \lambda + \mu \ket,
	\label{2.10}
\end{equation}
which says that $ \hat{P} $ is a generator of translation along $ S^1 $.
It should be noticed that an extra phase factor $ e^{ - i \, \alpha \, \mu } $
is multiplied.
These states
$ | \lambda \ket \, ( 0 \le \lambda < 2 \pi ) $
define a wave function $ \psi ( \lambda ) $
for an arbitrary state $ | \psi \ket \in H_\alpha $
by $ \psi ( \lambda ) := \bra \lambda | \psi \ket $.
This definition gives an isomorphism between $ H_\alpha $
and $ L^2 ( S^1 ) $ that is a space of square-integrable functions on $ S^1 $.
A bit calculation shows that the operators act on the wave function as
\begin{eqnarray}
	&&
	\hat{U} \psi ( \lambda )
	:= \bra \lambda | \hat{U} | \psi \ket
	= e^{ i \, \lambda } \, \psi ( \lambda ),
	\label{2.11}
	\\
	&&
	\hat{P} \psi ( \lambda )
	:= \bra \lambda | \hat{P} | \psi \ket
	= \left( - i \frac{\partial}{\partial \lambda} + \alpha \right)
	\psi ( \lambda ).
	\label{2.12}
\end{eqnarray}
In the last expression the parameter $ \alpha $ is interpreted
as the vector potential
for magnetic flux $ \Phi = 2 \pi \alpha $ surrounded by $ S^1 $.
Physical significance of $ \alpha $ is further discussed
in the reference~\cite{Tani}.
%
\section{Algebra}
\subsection{Fundamental algebra} 
Next we would like to extend Ohnuki-Kitakado's quantum mechanics on $ S^1 $
to a field-theoretical model.
We shall propose an algebra of the model.
To motivate definition of the algebra
we remind another expression of the canonical commutation relations.
If we put $ \hat{V}(a) := \exp ( - i \sum_j \, a_j \hat{p}_j ) $
for real numbers $ a = ( a_1 , \cdots , a_n ) $,
$ \hat{V}(a) $ is a unitary operator and satisfies
\begin{eqnarray}
	&&
	\hat{x}_j \, \hat{x}_k = \hat{x}_k \, \hat{x}_j,
	\label{3.1}
	\\
	&&
	\hat{V}(a)^\dagger \, \hat{x}_j \, \hat{V}(a) = \hat{x}_j + a_j,
	\label{3.2}
	\\
	&&
	\hat{V}(a) \, \hat{V}(b) = \hat{V}( a + b ).
	\label{3.3}
\end{eqnarray}
Geometrical meaning of the above algebra is obvious;
positions $ \hat{x} $'s are simultaneously measurable
and movable by the displacement operator $ \hat{V}(a) $;
displacement operators satisfy associativity.
\par
Quantum mechanics on $ S^1 $ is easily generalized
to the one on $ n $-dimensional torus $ T^{\, n} = ( S^1 )^n $.
 For this purpose we introduce
unitary operators $ \hat{U}_j $ and
self-adjoint operators $ \hat{P}_j ( j = 1 , \cdots , n ) $.
Put $ \hat{V}( \mu ) := \exp( -i \sum_j \, \mu_j \hat{P}_j ) $
for $ \mu = ( \mu_1 , \cdots , \mu_n ) \in \R^n $.
Naive generalization of (\ref{2.1}) leads the following relations;
\begin{eqnarray}
	&&
	\hat{U}_j \, \hat{U}_k = \hat{U}_k \, \hat{U}_j,
	\label{3.4}
	\\
	&&
	\hat{V}(\mu)^\dagger \, \hat{U}_j \, \hat{V}(\mu)
	= e^{ i \mu_j } \, \hat{U}_j,
	\label{3.5}
	\\
	&&
	\hat{V}(\mu) \, \hat{V}(\nu) = \hat{V}( \mu + \nu ).
	\label{3.6}
\end{eqnarray}
Representations of this algebra are constructed
by tensor products of Ohnuki-Kitakado representations
$ H_{\alpha_1} \otimes \cdots \otimes H_{\alpha_n} $.
Therefore irreducible representations are parametrized by $ n $-tuple parameter
$ \alpha = ( \alpha_1 , \cdots , \alpha_n ) $.
\par
Now we turn to the abelian sigma model in $ (1+1) $ dimensions.
The space-time is $ S^1 \times \R $
and the target space is $ S^1 $, on which the group $ U(1) $ acts.
In the classical sense
the model has a field variable $ \phi \in Q = \map(S^1; S^1) $.
Let $ \Gamma = \map(S^1; U(1)) $ a group by pointwise multiplication.
The group $ \Gamma $ acts on the configuration space $ Q $ by pointwise action;
for $ \gamma \in \Gamma $ and $ \phi \in Q $
let us define $ \gamma \cdot \phi \in Q $ by
\begin{equation}
	( \gamma \cdot \phi ) ( \theta ) := \gamma(\theta) \cdot \phi(\theta)
	\;\;\;
	(\theta \in S^1),
	\label{3.7}
\end{equation}
where $ \theta $ denotes a point of the base space.
In the right-hand side
the multiplication indicates the action of $ U(1) $ on $ S^1 $.
\par
To quantize this system let us assume that
$ \hat{\phi}(\theta) $ is a unitary operator for each point $ \theta \in S^1 $
and
$ \hat{V}(\gamma) $ is a unitary operator for each element
$ \gamma \in \Gamma $.
Moreover we assume the following algebra
\begin{eqnarray}
	&&
	\hat{\phi}(\theta)  \, \hat{\phi}(\theta') =
	\hat{\phi}(\theta') \, \hat{\phi}(\theta),
	\label{3.8}
	\\
	&&
	\hat{V}(\gamma)^\dagger \, \hat{\phi}(\theta) \, \hat{V}(\gamma) =
	\gamma(\theta) \, \hat{\phi}(\theta),
	\label{3.9}
	\\
	&&
	\hat{V}(\gamma) \, \hat{V}(\gamma') =
	e^{ - i \, c ( \gamma, \gamma' ) } \,  \hat{V}(\gamma \cdot \gamma')
	\;\;\;
	( \gamma, \, \gamma' \in \Gamma ).
	\label{3.10}
\end{eqnarray}
At the last line a function $ c : \Gamma \times \Gamma \to \R $
is called a central extension, which satisfies the cocycle condition
\begin{equation}
	  c( \gamma_1 , \gamma_2 )
	+ c( \gamma_1 \gamma_2 , \gamma_3 )
	= c( \gamma_1 , \gamma_2 \gamma_3 )
	+ c( \gamma_2 , \gamma_3 )
	\;\;\;
	( \mbox{mod} \: 2 \pi ).
	\label{3.11}
\end{equation}
If $ c \equiv 0 $, the algebra (\ref{3.8})-(\ref{3.10}) is
a straightforward generalization of (\ref{3.4})-(\ref{3.6})
to a system with infinite degrees of freedom.
We call the algebra (\ref{3.8})-(\ref{3.10})
the fundamental algebra of the abelian sigma model.
\par
To clarify geometrical implication of the algebra
we shall decompose the degrees of freedom of $ \phi \in Q $ and
$ \gamma \in \Gamma $.
In the classical sense we may rewrite $ \phi : S^1 \to S^1 \cong U(1) $ by
\begin{equation}
	\phi(\theta) = U \, e^{ i \, ( \varphi(\theta) + N \theta ) },
	\label{3.12}
\end{equation}
where $ U \in U(1) $, $ N \in \Z $ and
$ \varphi $ satisfies the no zero-mode condition;
\begin{equation}
	\map_0 (S^1; \R) :=
	\{
	\varphi : S^1 \to \R
	\, | \,
	C^\infty , \,
	\int_0^{2 \pi} \varphi(\theta) \, d \theta = 0
	\}.
	\label{3.13}
\end{equation}
The decomposition (\ref{3.12}) says that
$ Q \cong S^1 \times \map_0 ( S^1; \R ) \times \Z $.
Geometrical meaning of this decomposition is apparent;
$ U $ describes the zero-mode or collective motion of the field $ \phi $;
$ \varphi $ describes fluctuation or local degrees of freedom of $ \phi $;
$ N $ is nothing but the winding number.
Topologically nontrivial parts are $ U $ and $ N $.
\par
Similarly $ \gamma : S^1 \to U(1) $ is also rewritten as
\begin{equation}
	\gamma ( \theta ) = e^{ i \, ( \mu + f(\theta) + m \, \theta ) },
	\label{3.14}
\end{equation}
where $ \mu \in \R $, $ f \in \map_0 ( S^1 ; \R ) $ and $ m \in \Z $.
The action (\ref{3.7}) of $ \gamma $ (\ref{3.14}) on $ \phi $ (\ref{3.12})
is decomposed into
\begin{eqnarray}
	U & \to & e^{ i \mu } \, U,
	\label{3.15}
	\\
	\varphi(\theta) & \to & \varphi(\theta) + f(\theta),
	\label{3.16}
	\\
	N & \to & N + m.
	\label{3.17}
\end{eqnarray}
So the first component of $ \gamma $ (\ref{3.14}) translates the zero-mode;
the second one gives a homotopically trivial deformation;
the third one increases the winding number.
\par
As a nontrivial central extension for $ \gamma $ (\ref{3.14}) and
\begin{equation}
	\gamma' ( \theta ) = e^{ i \, ( \nu + g(\theta) + n \, \theta ) },
	\label{3.18}
\end{equation}
we define
\begin{equation}
	c( \gamma , \gamma' ) :=
	\frac{1}{4 \pi} \int_0^{2 \pi}
	\left( f'(\theta) g(\theta) - f(\theta) g'(\theta) \right) d \theta
	+ m \, \nu
	- n \, \mu.
	\label{3.19}
\end{equation}
This central extension is the simplest but nontrivial one
which is invariant under the action of $ \diff( S^1 ) $;
$ c( \gamma \circ \omega , \gamma' \circ \omega ) = c( \gamma , \gamma' ) $
for any $ \omega \in \diff( S^1 ) $.
The group $ \Gamma $ associated with such an invariant central extension
is called a Kac-Moody group.
The relation (\ref{3.10}) means that
$ \hat{V} $ is a unitary representation of the Kac-Moody group.
 For classification of central extensions see the literature \cite{Segal}.
%
\subsection{Algebra without central extension}
According to decomposition of classical variables
(\ref{3.12}) and (\ref{3.14}), quantum operators are also to be decomposed.
 For simplicity we consider the fundamental algebra (\ref{3.8})-(\ref{3.10})
without the central extension, that is, here we restrict $ c \equiv 0 $.
\par
Corresponding to (\ref{3.12}) we introduce
a unitary operator $ \hat{U} $,
self-adjoint operators\footnote{%
Expressing rigorously $ \hat{\varphi} (\theta) $
is an operator-valued distribution.}
$ \hat{\varphi} ( \theta ) $ for each $ \theta \in S^1 $
constrained by
\begin{equation}
	\int_0^{ 2 \pi } \hat{\varphi} ( \theta ) \, d \theta = 0,
	\label{3.20}
\end{equation}
and a self-adjoint operator $ \hat{N} $ satisfying
\begin{equation}
	\exp( 2 \pi i \, \hat{N} ) = 1,
	\label{3.21}
\end{equation}
which is called the integer condition for $ \hat{N} $.
We demand that the quantum field $ \hat{\phi} ( \theta ) $ is expressed
by them as
\begin{equation}
	\hat{\phi} ( \theta )
	=
	\hat{U} \, e^{ i \, ( \hat{\varphi} ( \theta ) + \hat{N} \theta ) }.
	\label{3.22}
\end{equation}
\par
Next, corresponding to (\ref{3.14}) we introduce
a self-adjoint operator $ \hat{P} $,
self-adjoint operators $ \hat{\pi} ( \theta ) $ for each $ \theta \in S^1 $
constrained by
\begin{equation}
	\int_0^{ 2 \pi } \hat{\pi} ( \theta ) \, d \theta = 0,
	\label{3.23}
\end{equation}
and a unitary operator $ \hat{W} $.
When $ \gamma $ is given by (\ref{3.14}), the operator $ \hat{V} ( \gamma ) $
is defined by
\begin{equation}
	\hat{V} ( \gamma )
	=
	e^{ - i \, \mu \, \hat{P} } \,
	\exp
	\left[
	- i \int_0^{2 \pi} f( \theta ) \, \hat{\pi} ( \theta )
	\, d \theta
	\right]
	\hat{W}^m.
	\label{3.24}
\end{equation}
\par
Using these operators the fundamental algebra is now rewritten as
\begin{eqnarray}
	&&
	[ \, \hat{P} , \hat{U} \, ] = \hat{U},
	\label{3.25}
	\\
	&&
	[ \, \hat{\varphi} ( \theta ) , \hat{\pi} ( \theta' ) \, ]
	=
	i \Bigl( \delta( \theta - \theta' ) - \frac{1}{2 \pi} \Bigr),
	\label{3.26}
	\\
	&&
	[ \, \hat{N} , \hat{W} \, ] = \hat{W},
	\label{3.27}
\end{eqnarray}
and all other commutators vanish.
In (\ref{3.26}) it is understood that the $ \delta $-function is defined
on $ S^1 $.
These commutators are equivalent to
\begin{eqnarray}
	&&
	e^{ i \, \mu \, \hat{P} } \, \hat{U} \, e^{ - i \, \mu \, \hat{P} }
	=
	e^{ i \, \mu } \, \hat{U},
	\label{3.28}
	\\
	&&
	\exp
	\left[
	  i \int_0^{2 \pi} f( \theta ) \hat{\pi} ( \theta )
	\, d \theta
	\right]
	\, \hat{\varphi} ( \theta ) \,
	\exp
	\left[
	- i \int_0^{2 \pi} f( \theta ) \hat{\pi} ( \theta )
	\, d \theta
	\right]
	=
	\hat{\varphi} ( \theta ) + f ( \theta ),
	\label{3.29}
	\\
	&&
	\hat{W}^\dagger \, \hat{N} \, \hat{W} = \hat{N} + 1.
	\label{3.30}
\end{eqnarray}
This algebra realize (\ref{3.15})-(\ref{3.17}) by means of (\ref{3.9}).
Observing the relation (\ref{3.30}) we call $ \hat{N} $ and $ \hat{W} $
the winding number and the winding operator, respectively.
Then remaining task is to construct representations of the algebra.
%
\subsection{Algebra with central extension}
Before constructing representations we reexpress the fundamental algebra
with the central extension (\ref{3.19})
respecting the decomposition (\ref{3.12}) and (\ref{3.14}).
The decomposition (\ref{3.22}) of $ \hat{\phi} $ still works.
On the other hand the decomposition (\ref{3.24}) of $ \hat{V} $
should be modified a little.
We formally introduce an operator $ \hat{\Omega} $ by
\begin{equation}
	\hat{W} = e^{ - i \, \hat{\Omega} }.
	\label{3.31}
\end{equation}
Although $ \hat{W} $ itself is well-defined,
$ \hat{\Omega} $ is ill-defined.
If $ \hat{\Omega} $ exists, (\ref{3.27}) would imply
$ [ \, \hat{N} , \hat{\Omega} \, ] = i $, 
which is nothing but the canonical commutation relation.
Therefore $ \hat{N} $ should have a continuous spectrum,
that contradicts the integer condition (\ref{3.21}).
Consequently $ \hat{\Omega} $ must be eliminated after calculation.
Bearing the above remark in mind, we replace (\ref{3.24}) by
\begin{equation}
	\hat{V} ( \gamma )
	=
	\exp
	\left[
		- i
		\Bigl(
			\mu \hat{P}
			+
			\int_0^{2 \pi} f( \theta ) \, \hat{\pi}(\theta )
			\, d \theta
			+
			m \, \hat{\Omega}
		\Bigr)
	\right].
	\label{3.32}
\end{equation}
 For the central extension (\ref{3.19})
it is verified that the following commutation relations should be added
to (\ref{3.25})-(\ref{3.27}) to satisfy the fundamental algebra;
\begin{eqnarray}
	&&
	[ \, \hat{\Omega} , \hat{P} \, ] = 2 i,
	\label{3.33}
	\\
	&&
	[ \, \hat{\pi} (\theta) , \hat{\pi} (\theta') \, ]
	=
	- \, \frac{i}{\pi} \, \delta' ( \theta - \theta' ).
	\label{3.34}
\end{eqnarray}
Using (\ref{3.31}) Eq. (\ref{3.33}) implies
\begin{equation}
	[ \, \hat{P} , \hat{W} \, ] = - 2 \, \hat{W},
	\label{3.35}
\end{equation}
which says that the zero-mode momentum $ \hat{P} $ is decreased by two units
when the winding number $ \hat{N} $ is increased by one unit
under the operation of $ \hat{W} $.
This is an inevitable consequence of the central extension.
We call this phenomenon ``twist''.
Using (\ref{3.33}) the decomposition (\ref{3.32}) results in
\begin{equation}
	\hat{V} ( \gamma )
	=
	e^{ - i \, \mu \, m }
	\,
	\exp
	\left[
		- i
		\Bigl(
			\mu \hat{P}
			+
			\int_0^{2 \pi} f( \theta ) \, \hat{\pi}(\theta )
			\, d \theta
		\Bigr)
	\right]
	\hat{W}^m .
	\label{3.37}
\end{equation}
\par
Here we summarize a temporal result;
with the notations (\ref{3.22}) and (\ref{3.37})
and the constraints (\ref{3.20}), (\ref{3.21}) and (\ref{3.23}),
the algebra
(\ref{3.25}), (\ref{3.26}), (\ref{3.27}), (\ref{3.34}) and (\ref{3.35})
is equivalent to
the fundamental algebra (\ref{3.8}), (\ref{3.9}) and (\ref{3.10})
including the central extension (\ref{3.19}).
Noticeable effects of the central extension are
the twist relation (\ref{3.35}) and the anomalous commutator (\ref{3.34}).
These features also affect representation of the algebra
as seen in the following sections.
%
\section{Representations}
%
\subsection{Without the central extension}
Now we proceed to construct representations of the algebra
defined by (\ref{3.25})-(\ref{3.27}) and other vanishing commutators
with the constraints (\ref{3.20}), (\ref{3.21}) and (\ref{3.23}).
\par
Remember that $ \hat{P} $ and $ \hat{N} $ are self-adjoint
and that $ \hat{U} $ and $ \hat{W} $ are unitary.
Both of the relations (\ref{3.25}) and (\ref{3.27})
are isomorphic to (\ref{2.1}).
Hence the Ohnuki-Kitakado representation works well for them.
$ \hat{P} $ and $ \hat{U} $ act on the Hilbert space $ H_\alpha $
via (\ref{2.3}) and (\ref{2.4}).
$ \hat{N} $ and $ \hat{W} $ act on another Hilbert space $ H_\beta $
via
\begin{eqnarray}
	&&
	\hat{N} \, | n + \beta \ket = ( n + \beta ) \, | n + \beta \ket,
	\label{4.1}
	\\
	&&
	\hat{W} \, | n + \beta \ket = | n + 1 + \beta \ket.
	\label{4.2}
\end{eqnarray}
The value of $ \alpha $ is arbitrary.
However $ \beta $ is restricted to be an integer
if we impose the condition (\ref{3.21}).
\par
 For $ \hat{\varphi} $ and $ \hat{\pi} $ the Fock representation works.
We define operators $ \hat{a}_n $ and $ \hat{a}^\dagger $ by
\begin{eqnarray}
	&&
	\hat{\varphi} (\theta)
	=
	\frac{1}{2 \pi} \, \sum_{ n \ne 0 } \,
	\sqrt{ \frac{ \pi }{ | n | } } \,
	( \hat{a}_n         \, e^{   i \, n \, \theta }
	+ \hat{a}_n^\dagger \, e^{ - i \, n \, \theta } ),
	\label{4.3}
	\\
	&&
	\hat{\pi} (\theta)
	=
	\frac{i}{2 \pi} \, \sum_{ n \ne 0 } \,
	\sqrt{ \pi | n | } \,
	( - \hat{a}_n         \, e^{   i \, n \, \theta }
	  + \hat{a}_n^\dagger \, e^{ - i \, n \, \theta } ).
	\label{4.4}
\end{eqnarray}
In the Fourier series the zero-mode $ n = 0 $ is excluded
because of the constraints (\ref{3.20}) and (\ref{3.23}).
It is easily verified that the commutator (\ref{3.26}) is equivalent to
\begin{equation}
	[ \, \hat{a}_m , \hat{a}_n^\dagger ] = \delta_{ m \, n }
	\;\;\;
	( m , n = \pm 1 , \pm 2 , \cdots )
	\label{4.5}
\end{equation}
with the other vanishing commutators.
Hence the ordinary Fock space $ F $ gives a representation
of $ \hat{a} $'s and $ \hat{a}^\dagger $'s.
\par
Consequently
the tensor product space $ H_\alpha \otimes F \otimes H_0 $ gives
an irreducible representation
of the fundamental algebra without the central extension.
The inequivalent ones are parametrized by $ \alpha $ $ ( 0 \le \alpha < 1 ) $.
\par
A remark is in order here;
the coefficients in front of $ \hat{a} $'s in (\ref{4.3}) and (\ref{4.4})
are chosen to diagonalize the Hamiltonian of free field
\begin{eqnarray}
	\hat{H}
	& := &
	\frac12
	\int_0^{ 2 \pi }
	\left[
	\Bigl( \frac{1}{ 2 \pi } \hat{P} + \hat{\pi}(\theta) \Bigr)^2
	+
	\Bigl( \partial \hat{\varphi}(\theta) + \hat{N} \Bigr)^2
	\right]
	d \theta
	\nonumber
	\\
	& = &
	\frac12
	\Bigl( \frac{1}{ 2 \pi } \hat{P}^2 + 2 \pi \hat{N}^2 \Bigr)
	+
	\sum_{ n \ne 0 } | n |
	\Bigl( \hat{a}_n^\dagger \, \hat{a}_n + \frac12 \Bigr).
	\label{4.6}
\end{eqnarray}
This Hamiltonian corresponds to the Lagrangian density
\begin{equation}
	{\cal L} =
	\frac{1}{2} \, \partial_\mu \phi^\dagger \, \partial^{\, \mu} \phi.
	\label{4.7}
\end{equation}
Interacting field theory will be briefly discussed later.
%
\subsection{With the central extension}
Next we shall construct representations of the algebra
defined by
(\ref{3.25}), (\ref{3.26}), (\ref{3.27}), (\ref{3.34}) and (\ref{3.35})
and other vanishing commutators
with the constraints (\ref{3.20}), (\ref{3.21}) and (\ref{3.23}).
The way of construction is similar to the previous one.
\par
Taking account of the twist relation (\ref{3.35}),
the representation of
$ \hat{P} $, $ \hat{U} $, $ \hat{N} $ and $ \hat{W} $ are given by
\begin{eqnarray}
	&&
	\hat{P}          \, | \, p + \alpha ; \, n \ket
	= ( p + \alpha ) \, | \, p + \alpha ; \, n \ket,
	\label{4.8}
	\\
	&&
	\hat{U} \, | \, p     + \alpha ; \, n \ket
	=          | \, p + 1 + \alpha ; \, n \ket,
	\label{4.9}
	\\
	&&
	\hat{N} \, | \, p + \alpha ; \, n \ket
	=    n     | \, p + \alpha ; \, n \ket,
	\label{4.10}
	\\
	&&
	\hat{W} \, | \, p     + \alpha ; \, n     \ket
	=          | \, p - 2 + \alpha ; \, n + 1 \ket.
	\label{4.11}
\end{eqnarray}
The inner product is defined by
\begin{equation}
	\bra p + \alpha ; \, m \, | \, q + \alpha ; \, n \ket
	=
	\delta_{ p \, q } \, \delta_{ m \, n }
	\;\;\;
	( p , q , m , n \in \Z ).
	\label{4.12}
\end{equation}
The Hilbert space formed by completing the space of linear combinations of
$ | \, p + \alpha ; \, n \ket $
is denoted by $ T_\alpha $.
($ T $ indicates ``twist''.)
\par
Let us turn to $ \hat{\varphi} $ and $ \hat{\pi} $.
Considering the anomalous commutator (\ref{3.34}),
after a tedious calculation we obtain a Fourier expansion
\begin{eqnarray}
	&&
	\hat{\varphi} (\theta)
	=
	\sum_{ n \ne 0 } \,
	\frac{1}{ \sqrt{ 2 | n | } } \,
	( \hat{a}_n         \, e^{   i \, n \, \theta }
	+ \hat{a}_n^\dagger \, e^{ - i \, n \, \theta } ),
	\label{4.13}
	\\
	&&
	\hat{\pi} (\theta)
	=
	\frac{i}{2 \pi} \, \sum_{ n = 1 }^{ \infty } \,
	\sqrt{ 2 n } \,
	( - \hat{a}_n         \, e^{   i \, n \, \theta }
	  + \hat{a}_n^\dagger \, e^{ - i \, n \, \theta } )
	\label{4.14}
\end{eqnarray}
and the commutation relations (\ref{4.5}).
It should be noticed that
only positive $ n $'s appear in the expansion of $ \hat{\pi} $
even though both of positive and negative $ n $'s appear in $ \hat{\varphi} $.
The algebra (\ref{4.5}) is also represented by the Fock space $ F $.
Hence the tensor product space $ T_\alpha \otimes F $ gives
an irreducible representation
of the fundamental algebra with the central extension
for each value of $ \alpha \, ( 0 \le \alpha < 1 ) $.
%
\section{Summary and discussion}
In this paper
we have defined the algebra of the abelian sigma model in $ (1+1) $ dimensions
and constructed representation spaces.
In the context of classical theory, this model has a field variable
$ \phi \in Q = \map( S^1 ; S^1 ) $.
The degrees of freedom are decomposed as
\begin{equation}
	Q \cong S^1 \times \map_0 ( S^1 ; \R ) \times \Z
	\label{5.1}
\end{equation}
by (\ref{3.12}).
The right-hand side is a direct product of topological spaces.
The first component represents the zero-mode;
the second one describes the fluctuation mode;
the third one corresponds the winding number.
Topological nature is concentrated in the first and the third components.
On the other hand
the group $ \Gamma = \map( S^1 ; U(1) ) $ acts on $ Q $ transitively.
Its covering group $ \tilde{\Gamma} $ is also decomposed as
\begin{equation}
	\tilde{\Gamma} \cong \R \times \map_0 ( S^1 ; \R ) \times \Z
	\label{5.2}
\end{equation}
by (\ref{3.14}).
The right-hand side is a direct product of topological groups.
We assign the algebra (\ref{3.8})-(\ref{3.10}) to $ Q $ and $ \tilde{\Gamma} $.
According to (\ref{5.1}) and (\ref{5.2}),
the algebra is decomposed into (\ref{3.25})-(\ref{3.27}).
When the central extension (\ref{3.19}) is included,
the anomalous commutator (\ref{3.34}) and the twist relation (\ref{3.35})
must be added.
An irreducible representation space is constructed by tensor product
of two Ohnuki-Kitakado representations with one Fock representation.
We obtain inequivalent ones parametrized by a parameter
$ \alpha \, ( 0 \le \alpha < 1 ) $.
The anomalous commutator eliminates negative-modes from $ \hat{\pi} (\theta) $
by means of (\ref{4.14}).
The twist relation causes (\ref{4.11});
the winding operator $ \hat{W} $ increases the winding number by one unit
and simultaneously decreases the zero-mode momentum by two units.
\par
Physical implication of our model is not clear yet.
Roles of the parameter $ \alpha $, the anomalous commutator
and the twist relation are to be examined further.
As a model with interaction, the sine-Gordon model
\begin{equation}
	{\cal L} =
	\frac12 \partial_\mu \psi (x) \partial^{\, \mu} \psi (x)
	+
	\kappa^2 \, \cos ( \psi(x) )
	\label{5.3}
\end{equation}
may be interesting.
Substitution of $ \phi = e^{ i \, \psi} $ defines
the corresponding Hamiltonian
\begin{eqnarray}
	\hat{H}
	& := &
	\frac12
	\int_0^{ 2 \pi }
	\left[
	\Bigl( \frac{1}{ 2 \pi } \hat{P} + \hat{\pi} \Bigr)^2
	+
	\partial \hat{\phi}^\dagger \, \partial \hat{\phi}
	-
	\kappa^2 ( \hat{\phi} + \hat{\phi}^\dagger )
	\right]
	d \theta
	\nonumber
	\\
	& = &
	\frac12
	\Bigl( \frac{1}{ 2 \pi } \hat{P}^2 + 2 \pi \hat{N}^2 \Bigr)
	+
	\sum_{ n \ne 0 } | n |
	\Bigl( \hat{a}_n^\dagger \, \hat{a}_n + \frac12 \Bigr)
	\nonumber
	\\
	&&
	-
	\frac{\kappa^2}{2}
	\int_0^{ 2 \pi } \Bigl(
		\hat{U} e^{i \, ( \hat{\varphi} ( \theta ) + \hat{N} \theta ) }
		+
		\hat{U}^\dagger
		e^{ - i \, ( \hat{\varphi} ( \theta ) + \hat{N} \theta ) }
	\Bigr) d \theta.
	\label{5.4}
\end{eqnarray}
The last term yields highly nonlinear complicated interaction.
It is known \cite{Coleman} that
this model has a topological soliton, which behaves like a fermion.
It is expected that our formulation may shed light on the soliton physics
of sigma models.
\par
 From both points of view, field theory and string theory,
it is hoped to extend our model
to nonabelian cases and to higher dimensions.
Our model has a field configuration manifold $ Q = \map( S^1 ; S^1 ) $.
The most general model has $ Q = \map( M ; N ) $.
(1)An immediate extension is
a choice $ M = S^1 $ and $ N = T^{\, n} = ( S^1 )^n $.
This is the toroidal compactification model of string theory \cite{Narain}.
(2)A rather easy extension is
a choice $ M = S^n $ or $ T^{\, n} $ and $ N = S^1 $.
 For $ M = S^n ( n \ge 2 ) $ there is no winding number.
Yet we should pay attention to the zero-mode.
There may be nontrivial central extensions.
(3)Another nontrivial extension is
a choice $ M = S^n $ or $ T^{\, n} $ and $ N = G $, that is a Lie group.
This corresponds to a chiral Lagrangian model.
 For a nonabelian group $ G $, we know neither existence nor uniqueness
of the decomposition
\begin{equation}
	\Gamma
	:= \map( S^n ; G )
	\cong G \times \map_0 ( S^n ; \mbox{Lie}(G) ) \times \pi_n(G),
	\label{5.5}
\end{equation}
where $ \pi_n $ denotes the $ n $-th homotopy group and
\begin{equation}
	\map_0 ( S^n ; \mbox{Lie}(G) )
	:=
	\{
		\, g : S^1 \to \mbox{Lie}(G)
		\, | \,
		C^\infty, \,
		\int_0^{2 \pi} g ( \theta ) = 0
	\}.
	\label{5.6}
\end{equation}
Even if it exists, it may not be a direct product of topological groups
because of nonabelian nature.
(4)A highly nontrivial one is
a choice $ N = G/H $, that is a homogeneous space.
This model is a nonlinear sigma model.
Quantum mechanics on $ G/H $ is already well-established \cite{Tani2}.
However extension to field theory remains difficult.
(5)Another interesting one is
a choice $ M = S^1 $ and $ N = T^{\, n}/P $, that is an orbifold.
This is nothing but the orbifold model of string theory \cite{Sakamoto}.
It is found \cite{Sakamoto} that
zero-mode variables obey peculiar commutators
and nontrivial quantization is obtained
for string theory on an orbifold with a background 2-form.
It is expected that our model may work as a simplified model to understand
such a complicated behavior.
%
\section*{Acknowledgments}
This investigation is stimulated by seminars given
by Prof. Sakamoto and Dr. Tachibana.
%

%
\end{document}